\documentclass[aps,pra,showpacs,twocolumn,eqsecnum]{revtex4}
\usepackage[dvips,final]{graphicx}
\usepackage{bm, graphics, amsmath}
\begin{document}

\title{Optimal quantum state identification with qudit-encoded unknown  states}
\author{Ulrike Herzog}
\affiliation{Nano-Optics Group, Institut f\"ur Physik, Humboldt-Universit\"at zu Berlin, Newtonstrasse 15, D-12489 Berlin, Germany}

\date{\today}

\begin{abstract}
We consider the problem of optimally identifying the state of a probe qudit,
prepared with given prior probability in a pure state belonging to a finite set of possible states which together span a $D$-dimensional subspace of the $d$-dimensional Hilbert space the qudit is defined in. 
It is assumed that we do not know some or all of the states in the set, but for each unknown state we are given a reference qudit into which this state is encoded. 
We show that from the measurement for optimal state identification with  $d=D$ one can readily determine the optimal figure of merit for qudits with  $d>D$, without solving a new optimization problem. This result is applied to the  minimum-error identification and to the optimal unambiguous identification of two qudit states with $d\geq 2$, where either one or both of the states are unknown,
and also to the optimal unambiguous identification of $N$ equiprobable linearly independent unknown pure qudit states with $d \geq  N$. 
In all cases the optimal figure of merit,  averaged over the unknown states, increases with growing dimensionality $d$ of the qudits.     
 
\end{abstract}


\maketitle

\section{Introduction}

In quantum state identification \cite{BH, hayashi1, hayashi2, BBFHH, zhang, HeB2,IHHH, HB-d, sentis1, akimoto, colin, sunian, sentis2, zhou}
 the task is to identify the state of a quantum system,  prepared with certain prior probability in a definite state out of a finite  set of possible states, where some or all of the states are unknown. The unknown states are encoded into different reference copies, which together have been introduced as the program register in a programmable machine \cite{BH} for identifying the state of the probe, carrying the data. We assume that the states to be identified are qudits, defined  in a $d$-dimensional Hilbert space. Identifying an unknown probe  state means to find the particular reference copy whose state matches the state of the probe. 
 For this purpose a measurement  for quantum state discrimination \cite{chefles, springer, barnett-croke} has to be performed on the combined system, composed of the  probe  qudit and the reference qudits, where due to the averaging with respect to the unknown states 
the combined system is described by a mixed state.   
Because of the specific symmetry properties inherent in the different states of the combined system, 
they can be discriminated in spite of the complete lack of knowledge about the unknown states. 

In general the states of the combined system are not orthogonal, which implies that their discrimination and hence the  identification of the state of the probe cannot be perfect. 
For discriminating nonorthogonal states a number of 
optimal strategies have been developed, which are optimized with respect to various figures of merits. The best known
of these strategies are discrimination with minimum error \cite{helstrom,holevo}, where every time a measurement is performed a decision about the state is made,  and optimal
unambiguous discrimination \cite{ivanovic,dieks,peres,jaeger}, where errors
are not allowed, at the expense of admitting inconclusive
results the probability of which is minimized. 

 The problem of quantum state identification has been first introduced  for the optimal unambiguous identification of two unknown pure qubit states  \cite{BH}.    
The investigations have  been soon  extended  to minimum-error identification \cite{hayashi1}, and generalizations have been also performed to take into account that more than one copy may be available  for the probe system and the two reference systems \cite{hayashi1, hayashi2, IHHH, BBFHH, HeB2, sentis1,akimoto,zhou}, including the case that the two unknown states are mixed \cite{sentis1}. 
For two unknown pure qudit states with Hilbert space dimension $ d \geq 2$,  minimum-error identification has been studied for arbitrary prior probabilities of the states \cite{IHHH}, while the maximum overall success probability for unambiguous identification has been obtained in the case when the states are equally probable  \cite{hayashi2,IHHH}.  
A variant of the state identification problem, where for $d=2$ one of the two  pure states is known and the other is unknown, has also been studied  \cite{BBFHH}. 
In addition, for two unknown qubit states  the case where  some classical knowledge is available has been  treated \cite{colin, sunian}, and a modified identification strategy with a fixed error rate has been investigated \cite{sentis2}. 

Only very few results have been obtained so far for optimal state identification with more than two unknown states.  For $d$ linearly independent unknown pure qudit states  a strategy yielding the worst-case success probability for the  unambiguous identification  without knowledge of the prior probabilities of the states has been studied \cite{zhang}. Supposing that these states have equal prior probabilities, the measurement for their optimal  unambiguous identification, yielding the maximum overall success probability, has been  derived in our earlier paper \cite{HB-d}. 

In the present paper we address the question as to how  the optimal figure of merit for qudit state identification depends on the qudit dimensionality $d.$ We assume that we know the prior probabilities of the states in the given set but that some or all of them are unknown to us.  
Supposing that  the possible qudit states  span a $D$-dimensional Hilbert space, we  show in Sec. II that from an optimal measurement for state identification with $d=D$ one can readily obtain the optimal figure of merit for qudits with  $d>D$, without solving a new optimization problem, that is without explicitly determining the optimal operators characterizing the  measurement for $d>D.$ In Sec. III we apply this result to the  minimum-error identification and to the optimal unambiguous identification of two qudit states with $d\geq 2$, occurring with arbitrary prior probabilities, where either one or both of the states are unknown, and also to the optimal unambiguous identification of $N$ equiprobable linearly independent unknown pure qudit states with $d\geq N$. The paper is concluded in Sec. IV.

\section{General theory}

\subsection{Formulation of the problem}

Our task is to  identify the state of a  $d$-dimensional quantum system, the probe qudit, which is prepared with given prior probabilities $\eta_1,\ldots,\eta_{N^{\prime}}$  in one of $N^{\prime}$  pure states $ |\psi_{1}\rangle, \ldots,  |\psi_{N}\rangle,\ldots,|\psi_{N^{\prime}}\rangle$. The $N$ states  $ |\psi_{1}\rangle, \ldots, |\psi_{N}\rangle$ are unknown to us, but instead we possess $N$  reference  systems of the same kind as the probe,  into which the $N$ unknown states  are encoded. 
Our  total quantum system thus  consists of one probe qudit, labeled by the index zero,  and $N$
reference qudits,  labeled by the indices $1, \ldots, N$. 
 Let the probe qudit  be in the  state $|\psi_n\rangle_0$. If all possible states were known, the  combined $(N+1)$-qudit system would be described by the state vector $|\Psi_{n}\rangle $ with  
\begin{equation}
\label{Psi}
|\Psi_{n}\rangle = |\psi_{n}\rangle_0 \;|\psi_{1}\rangle_1 \ldots  |\psi_{N}\rangle_N \quad (n=1,\ldots, N, \ldots,  N^{\prime}),
\end{equation}
where the tensor-product signs have been omitted. Since the states $ |\psi_{1}\rangle, \ldots, |\psi_{N}\rangle$ are unknown,   
the average  has to be taken with respect to these states. After a suitable parametrization,  the averaging procedure  amounts to multiple   
integrations in the  parameter space $\Gamma^{(d)}$ corresponding to the $N$ unknown  qudit states. 
Hence the possible states of the combined  quantum system are described by the density operators    
\begin{equation}
\label{av}
 \rho_n^{(d)}=  \int \!{\rm d}\Gamma ^{(d)} |\Psi_{n} \rangle \langle\Psi_{n}| \quad {\rm with}\;\;{\rm d}\Gamma ^{(d)}={\rm d}\Gamma_1 ^{(d)} \ldots {\rm d} \Gamma_N ^{(d)}
\end{equation}
for  $n=1, \ldots  N^{\prime}$. Here  
${\rm d}\Gamma_j^{(d)}$ refers to the  parameter space of  an  unknown qudit state $ |\psi_j\rangle$ and we suppose the normalization  condition  $ \int \!{\rm d}\Gamma_j ^{(d)}\cdot 1=1.$ 
The qudit states can be expanded as $|\psi_{j}\rangle=\sum_{i=0}^{d-1}c_{ji}|i\rangle$  where $\{|i\rangle \}$  denotes an arbitrary orthonormal basis.  The unknown states are assumed to be independently and randomly
chosen from their state space, that is the $d$ complex expansion coefficients $c_{ji}$ are uniformly distributed  under the constraint $\sum_{i=0}^{d-1}|c_{ji}|^2 =1$ where  the distribution does not depend on the choice of
the basis. 
In particular, this means that  
\begin{equation}
\label{rel}
\int \!{\rm d}\Gamma_j^{(d)}  |\psi_{j}\rangle_l \langle\psi_{j}|_l  =\frac{I_l^{(d) }}{d} \quad{\rm with} \;\;I_l^{(d)}=\sum_{i=0}^{d-1} |i\rangle_l \langle i|_l.
\end{equation}
From Eqs. (\ref{Psi}) and (\ref{av}) we obtain  in analogy to Ref. \cite {HB-d}
\begin{equation}
\label{rho_n}
\rho_n^{(d)}=\begin{cases}
 \dfrac{2}{(d+1)d^N}\;P_{0,n}^{sym(d)} \bigotimes \limits_{l=1\atop
l\neq n}^N I_l^{(d)} &\text{if }1\leq n \leq N \\ 
\dfrac{1}{d^N}\, |\psi_n\rangle_0\langle\psi_n|_0 \bigotimes \limits_{l=1}^N I_l^{(d)}
&\text{if } N <n \leq  N^{\prime}.  
\end{cases}
\end{equation}
 Here  the first line refers to the case that  the probe state $|\psi_n\rangle_0$ is unknown, while the second line applies when $|\psi_n\rangle_0$ is known and therefore does not need to be encoded into a reference copy.    
$P_{0,n}^{sym(d)}$ is the projector onto  
the   symmetric subspace of the two $d$-dimensional Hilbert spaces belonging to the probe qudit and to the $n$-th reference qudit, respectively. It  has the rank  $d(d+1)/2$ and can be represented as \cite {HB-d}
\begin{eqnarray}
\label{P0nsym}
 \lefteqn{P_{0,n}^{sym(d)} =\sum_{i=0}^{d-1} |i\rangle_0 |i\rangle_n \langle i|_0 \langle i|_n}
 \\
&&+\sum_{j=1}^{d-1} \sum_{i=0}^{j-1} \frac{|i\rangle_0 |j\rangle_n + |j\rangle_0 |i\rangle_n}{\sqrt{2}} \;\frac{
\langle i|_0 \langle j|_n +  \langle j|_0 \langle i|_n}{\sqrt{2}}, 
 \nonumber
\end{eqnarray}
where the orthonormal basis states $|0\rangle,\ldots|d-1\rangle$ are arbitrary.  
 For later use we  note that    
\begin{equation}
\label{rel3}
\int \!{\rm d}\Gamma_j^{(d)} |\psi_{j}\rangle_0   |\psi_{j}\rangle_n  \langle\psi_{j}|_0 \langle\psi_{j}|_n =\frac{2}{d(d+1)}
P_{0,n}^{sym(d)},
\end{equation}
as becomes obvious from the first line of Eq. (\ref{rho_n}) together with Eqs. (\ref{Psi}) - (\ref{rel}). 

In order to identify the state of the probe qudit  we have to discriminate  between the $N^{\prime}$
nonorthogonal  mixed states of our combined quantum system, given by Eq. (\ref{av}) or, equivalently, by  Eq. (\ref{rho_n}).  
 A measurement  discriminating between $N^{\prime}$ states is in general   
described by  $N^{\prime}+1$ positive detection operators
$\Pi_0, \Pi_1, \ldots \Pi_{N^{\prime}},$  where ${\rm Tr}({\rho_{n}}{\Pi}_{0})$  yields the probability  that the
  result obtained in the presence of the   $n$-th state   is inconclusive, while ${\rm
Tr}({\rho_n}{\Pi}_n)$ and   ${\rm Tr}({\rho_n}{\Pi}_m)$  with $m\neq 0,n$ are the probabilities for obtaining a correct and erroneous result, respectively \cite{chefles,springer,barnett-croke}.  
The detection operators  fulfill the completeness relation  
\begin{equation}
\label{compl}
\Pi_0 +  \sum_{n=1}^{N^{\prime}}  \Pi_n = I=\bigotimes \limits_{l=0}^N I_l^{(d)}=\sum_{l=0}^{N}\sum_{i=0}^{d-1} |i\rangle_l \langle i|_l,
\end{equation}
where $I$ is the identity operator in the   $d^{N+1}$-dimensional Hilbert space $ {\cal H}$ belonging to our combined  quantum system, composed of the probe qudit and the $N$ reference qudits. Since the prior probabilities of the possible states obey the relation  $\sum_{n=1}^{N^{\prime}}\eta_n = 1$, the overall probability $P_c$ of correctly identifying the state of the probe qudit is given by 
\begin{equation}
P_c=  \sum_{n=1}^{N^{\prime}}\eta_n {\rm Tr}\left(\rho_n^{(d)} \Pi_n\right). 
\label{Pcorr}
\end{equation}

In this paper it is our aim to find the maximum value of  $P_{c}$, subject to the constraints imposed by the two best-known  optimal strategies  \cite{chefles,springer,barnett-croke}  for discriminating the  states  $\rho_n^{(d)}.$ 
In the strategy of minimum-error discrimination  it is required that inconclusive results do not occur, which leads to the constraint $\Pi_0=0$. We then get $P_E = 1- P_C$, where $P_E$ is the minimum overall error probability and where we used the notation $P_C$ for the maximum overall probability of correct discrimination,
\begin{equation}
  P_c^{max} = P_C \qquad {\rm  if}\;\;\Pi_0 =I-\sum_{n=1}^{N^{\prime}}  \Pi_n = 0.
\label{minerr}
\end{equation}
On the other hand, in the strategy of optimal unambiguous discrimination errors do not occur. This can be achieved probabilistically, at the expense of admitting a certain minimum probability  $Q_F$ that the measurement fails to give a conclusive outcome.  The absence of errors means that ${\rm Tr}(\rho_n {\Pi}_m)=0$ for $m\neq n,0,$ which requires that   
${\rho_n}{\Pi}_m=0,$  due  to the positivity of the operators $\rho_n$ and ${\Pi}_m$. It follows that $Q_F=1-P_S$, where $P_S$ denotes the maximum probability that the measurement succeeds to give a conclusive result, which in this strategy is always correct. Hence we arrive at the maximum success probability     
\begin{equation}
P_{c}^{max}=P_S   \qquad {\rm if}\;\;\Pi_0 \geq 0,\;\;{\rho_n}{\Pi}_m=0\;( m\neq n),
\label{unamb}
\end{equation}
where  $m,n=1,\dots, N^{\prime}.$ The two different abbreviations  chosen for $P_{c}^{max}$ reflect the different constraints 
applied in the two discrimination strategies.

\subsection{Treatment for arbitrary qudit-dimensionality  $ d$   } 

Let $D$ denote the dimensionality of the Hilbert space jointly spanned by the $N^{\prime}$ pure states to be identified, where $D<N^{\prime}$ for linearly dependent states, while  $D=N^{\prime}$  when the input states are linearly independent.  
Since $D$ cannot be larger than the dimensionality $d$ of the qudits into which the states are encoded, it follows that $d\geq D$. The operators $\Pi_n$ characterizing the optimal measurement for identifying the states are obtained  when the probability $P_c$ given in Eq. (\ref{Pcorr}) is maximized, subject to the constraints imposed by the specific measurement strategy, given in Eqs. (\ref{minerr}) or   (\ref{unamb}), respectively. The density operators  occurring in Eq. (\ref{Pcorr}) are described by Eq. (\ref{rho_n}) and do not depend on the unknown states, but solely rely on symmetry properties. The determination of the optimal operators  $\Pi_n$  requires the solution of an optimization problem in the full  $d^{N+1}$-dimensional Hilbert space $ {\cal H}.$ 

In the following we show that for $d>D$ the value of  the maximum probability $ P_c^{max}$ can be also derived   from the solution of the corresponding optimization problem in a Hilbert space with the reduced dimension $D^{N+1},$  and we investigate how this value depends on $d$ when $D$ is fixed. For this purpose we define for each of our $N+1$ qudits the Hilbert space 
\begin{equation}
\label{D} 
{\cal H}_l^{\Psi} \! = 
{\rm span}   \{ |\psi_{1}\rangle_l, \ldots, |\psi_{N^{\prime}}\rangle_l\}\quad {\rm with}\;\;{\rm dim}\{{\cal H}_l^{\Psi}\}=D, 
\end{equation}
where again $l=0$ refers to the probe qudit and $l=1, \ldots,N$ to the $N$ reference qudits.  ${\cal H}_l^{\Psi}$ is a $D$-dimensional subspace of the $d$-dimensional Hilbert space   belonging to the $l$th qudit. By applying the Hilbert-Schmidt orthonormalization procedure to the input states, we can construct  an  orthonormal basis in ${\cal H}_l^\Psi.$ Numbering the input states in such a way that the first $D$ of them are linearly independent, we introduce the  set of $D$ orthonormal basis states $\{|j\rangle\}$ with    $j= 0, \ldots,  D-1,$  defined as 
\begin{eqnarray}
 \label{basis-a}
&&|0\rangle= |\psi_1\rangle,\quad |1\rangle=\frac{|\psi_2\rangle-  |\psi_1\rangle\langle \psi_1|\psi_2\rangle}{\sqrt{1-|\langle \psi_1|\psi_2\rangle |^2}},\\
 \label{basis-b}
&& |j\rangle =\frac{1}{\sqrt{{\cal N}_j}}
\left(|\psi_{j+1}\rangle-\sum_{k=0}^{j-1}| k \rangle \langle k |\psi_{j+1}\rangle\right)\quad \\
 \label{basis-c}
&&{\rm with}\;\; {\cal N}_j = 1- \sum_{k=0}^{j-1} |\langle \psi_{j+1}|k\rangle|^2 = |\langle \psi_{j+1}|j\rangle|^2,  \qquad
\end{eqnarray}
where the subscripts $l$ have been omitted.  
Here the  second equality sign in Eq.  (\ref{basis-c}) follows directly from Eq. (\ref{basis-b}). 
Taking into account that  with these  basis states Eq. (\ref{D}) can be alternatively written as    
${\cal H}_l^{\Psi} \!\! = {\rm span}   \{ |0\rangle_l, \ldots, |D-1 \rangle_l\}$,
we introduce the projector   
\begin{equation}
\label{P} 
\hat{P}_{\Psi} = 
\sum_{l=0}^N\sum_{j=0}^{D-1} |{j}\rangle_l  \,\langle {j}|_l, \quad 
{\rm where} \,\,\;  \hat{P}_{\Psi}|\Psi_n\rangle = |\Psi_n\rangle.
\end{equation}
 $\hat{P}_{\Psi}$ projects onto the $D^{N+1}$-dimensional Hilbert space  
\begin{equation}
\label{H-psi} 
 {\cal H}_{\Psi} = 
{\cal H}_0^{\Psi}\otimes\ldots \otimes {\cal H}_N^{\Psi},
\end{equation}
which is a subspace of our total $d^{N+1}$-dimensional Hilbert space  ${\cal H}.$ 
 Eqs. (\ref{basis-a})  - (\ref{P}) define the projector $\hat{P}_{\Psi}$ for an arbitrary set of linearly independent   
input states  $|\psi_1\rangle,\ldots, |\psi_D\rangle$. We suppose that these states  include the unknown states, which means that the projector $\hat{P}_{\Psi}$ itself is also unknown. Since  $\hat{P}_{\Psi}$ is constructed in such a way that the right equality in Eq. (\ref{P}) holds for arbitrary input states,  no matter whether they  are known or unknown, it follows from Eq. (\ref{av}) that      
\begin{equation}
\label{trace}
{\rm Tr}\left(\rho_n ^{(d)}\Pi_n\right) \!=\!
\int \!\!{\rm d}\Gamma^{(d)}   \langle\Psi_{n}| \Pi_n  |\Psi_{n}\rangle \!= \! \int\! \!{\rm d}\Gamma^{(d)}   \langle\Psi_{n}| \Pi_n^{\Psi}  |\Psi_{n}\rangle,  \\
\end{equation}
where we introduced the operator   
\begin{equation}
\label{pos}
{\Pi}_n^{\Psi}= \hat{P}_{\Psi} \Pi_n \hat{P}_{\Psi} \quad {\rm with}\;\; \sum_{n=0}^{N^{\prime}} {\Pi}_n^{\Psi} = \hat{P}_{\Psi}.
\end{equation}
 The overall  probability of correctly identifying the state of the probe qudit, given by Eq. (\ref{Pcorr}), then takes the alternative form 
\begin{equation}
P_c= 
 \sum_{n=1}^{N^{\prime}}\eta_n \!\int\! \!{\rm d}\Gamma^{(d)}   \langle\Psi_{n}| \Pi_n^{\Psi}  |\Psi_{n}\rangle.
\label{Pcorr1}
\end{equation}
Eq. (\ref{Pcorr1}) shows that we can obtain the maximum of  $P_c$ by first determining the optimal operators $\Pi_n^{\Psi}$ $(n=1,\ldots N^{\prime}$), that is by expressing these operators in terms  of the states  $|0\rangle, \ldots,|D-1\rangle$ which according to Eqs. (\ref{basis-a}) and  (\ref{basis-b}) depend on the input states, and by subsequently performing  the integrations with respect to the parameter spaces of the unknown states. 
Since the optimal operators  $\Pi_n^{\Psi},$ yielding after integration the maximum of $P_c,$ act in the $D^{N+1}$-dimensional subspace ${\cal H}_{\Psi},$ they have to  maximize the overall probability of  getting a correct result given that the mixed state of the combined system falls into this  particular subspace. This means that the optimal operators $\Pi_n^{\Psi}$ also maximize the overall  joint probability $P_c^{\Psi}$ that the identification is correct and  the state of the combined system falls into  ${\cal H}_{\Psi},$   
\begin{equation}
\label{Pcorr2}
P_c^{\Psi}=  
\sum_{n=1}^{N^{\prime}}\eta_n {\rm Tr}\left(\rho_n ^{\Psi}\Pi_n^{\Psi}\right) \quad{\rm with}\;\;
{\rho}_n^{\Psi}= \hat{P}_{\Psi} \rho_n^{(d)} \hat{P}_{\Psi}. 
\end{equation}
It should be noted that  ${\rm Tr}({\rho}_n ^{\Psi}) <1$ for $D< d,$  since ${\rm Tr}({\rho}_n ^{\Psi})$ describes the probability that the mixed state of the combined system falls into  the subspace ${\cal H}_{\Psi}$ when the probe qudit is prepared in the $n$-th state.  The overall probability that the  combined state is confined to ${\cal H}_{\Psi}$ is given by  $\sum_{n=1}^{N^{\prime}}\eta_n {\rm Tr}(\rho_n ^{\Psi}).$  We emphasize that  $P_c^{\Psi}$ differs from $P_c$ unless ${\cal H}_{\Psi}$ is identical with ${\cal H},$ that is unless  $D=d.$  

The crucial point for determining $\rho_n ^{\Psi}$  is the fact that the representation of $\rho_n^{(d)},$ given by  Eq. (\ref{rho_n}),
 holds true when an arbitrary  $d$-dimensional basis is used in the Hilbert spaces belonging to the individual qudits. Hence  the  $d$-dimensional basis  can always be chosen in such a way that the first $D$  basis states are given by   Eqs.  (\ref{basis-a}) -  (\ref{basis-c}). The operators  ${\rho}_n^{\Psi}= \hat{P}_{\Psi} \rho_n^{(d)} \hat{P}_{\Psi}$  then take the explicit form     
\begin{equation}
\label{rho-n-Psi}
\rho_n^{\Psi}=\begin{cases}
 \dfrac{2}{(d+1)d^N}\;P_{0,n}^{sym({\Psi})} \bigotimes \limits_{l=1\atop
l\neq n}^N I_l^{(\Psi)} &\text{if }1\leq n \leq N \\ 
\dfrac{1}{d^N}  \hat{P}_{\Psi}|\psi_n\rangle_0\langle\psi_n|_0  \hat{P}_{\Psi}\;\bigotimes \limits_{l=1}^N I_l^{(\Psi)}
&\text{if } N <n \leq  N^{\prime},  
\end{cases}
\end{equation}
where  $ I_l^{(\Psi)}$ and $P_{0,n}^{sym(\Psi)}$ are defined by  Eqs. (\ref{rel}) and (\ref{P0nsym}) with $d=D,$ and where the basis   states  occurring in these expressions are given   by Eqs.  (\ref{basis-a}) -  (\ref{basis-c}).   
The maximization of $P_c^{\Psi},$ yielding the optimal operators $\Pi_n^{\Psi}$, has to be performed on the constraints imposed by the specific strategies  used for the  optimal identification.
For minimum-error identification we have to require that    
\begin{equation}
\Pi_0^{\Psi} =\hat{P}_{\Psi}-\sum_{n=1}^{N^{\prime}}  \Pi_n^{\Psi} = 0,
\label{minerr1}
\end{equation}
 see  Eq. (\ref{minerr}), while  for optimal unambiguous identification the constraints   
\begin{equation}
\Pi_0^{\Psi} \geq 0, \quad  \rho_n^{\Psi} {\Pi}_m^{\Psi} =0\;\;( m\neq n,\; m,n=1,\dots, N^{\prime}).
\label{unamb1}
\end{equation}
 have to be satisfied, see Eq. (\ref{unamb}).

 Eqs. (\ref{Pcorr1}) - (\ref{unamb1}) together with Eqs.  (\ref{basis-a}) -  (\ref{basis-c}) are the main result of our paper. They reduce the original optimization problem, defined in the $d^{N+1}$-dimensional Hilbert space  ${\cal H},$  to an optimization problem in the  $D^{N+1}$-dimensional subspace ${\cal H}_{\Psi}$. The latter problem  is mathematically equivalent to the original problem with $d=D,$ as becomes obvious from Eqs. (\ref{Pcorr2}) and  (\ref{Pcorr}).  After maximizing $P_c^{\Psi}$ we arrive at the optimal  operators $\Pi_n^{\Psi}$, given in terms of the basis states defined by Eqs.  (\ref{basis-a}) -  (\ref{basis-c}).   
 The final result is obtained when the  expressions for   $\Pi_n^{\Psi}$ are substituted into  Eq. (\ref{Pcorr1}) and the integrations in the $d$-dimensional parameter spaces of the  unknown qudit states are carried out.
A similar treatment can be performed when instead  of the overall probability of correct results another figure of merit has to be optimized.  
Hence when  the possible qudit states span a $D$-dimensional Hilbert space,  from an optimal measurement for state identification with   $d=D$ one can  readily determine the optimal figure of merit for qudits with  $d>D,$ without solving a new optimization problem, and without explicitly determining  
the expressions for the  detection operators $\Pi_n$ that describe the measurement for optimal identification in the full $d^{N+1}$-dimensional Hilbert space  ${\cal H}.$  
 
Before proceeding, let us summarize  a few  equations which  will be needed for performing the integrations in Eq. (\ref{Pcorr1}). Omitting the subscript $l$, we find  from  Eq. (\ref{rel}) that   
\begin{equation}
\label{rel1}
\int \!{\rm d}\Gamma_j^{(d)} |\langle 0 |\psi_{j}\rangle |^2 =\frac{1}{d}, \quad \int \!{\rm d}\Gamma^{(d)}  |\langle \psi_1 |\psi_{2}\rangle |^2=\frac{1}{d},
\end{equation}
where $|0\rangle$ is an arbitrary state that does not depend on the unknown state  $|\psi_j\rangle.$ Here the second equation follows from the first since  $|\psi_1\rangle$ is independent of $|\psi_2\rangle$.  
Similarly,  using Eqs. (\ref{P0nsym}) and (\ref{rel3}) and      
 determining the expression 
 $\langle 0|_0 \langle 0|_n  P_{0,n}^{sym(d)}  |0\rangle_0   |0\rangle_n $ 
we arrive at 
\begin{eqnarray}
\label{rel4}
&&\int \!{\rm d}\Gamma_j^{(d)} |\langle 0 |\psi_{j}\rangle|^4 =\frac{2}{d(d+1)}.
\end{eqnarray}
From Eqs.  (\ref{basis-a}) and (\ref{basis-b}) it becomes obvious that for $j\leq D$ the input state $|\psi_{j}\rangle$ lies in the subspace spanned by the basis states $|0\rangle,\ldots, |j-1 \rangle$ and is therefore orthogonal to the remaining basis states $|{j}\rangle,\ldots, |{D-1}\rangle$. Hence it follows that  
\begin{equation}
\label{basis1}
 \langle \psi_{j}| k \rangle  = 0\quad  {\rm if} \;\; k \geq j.  
 \end{equation}
Moreover, since   the states   $|0\rangle,\ldots, |j-2\rangle$ do not depend on $|\psi_{j}\rangle $,  we obtain with the help of Eqs.  (\ref{basis-c}) and (\ref{rel1}) the relation  
\begin{equation}
\label{rel2}
\!\int \!\!{\rm d}\Gamma_{j}^{(d)} |\langle \psi_{j}|j\!-\!1\rangle|^2
=1- \sum_{k=0}^{j-2} \!\int \!\!{\rm d}\Gamma_{j}^{(d)}  | \langle \psi_{j}|k\rangle|^2
=1-\frac{j\!-\!1}{d}
\end{equation}
for $j=1,\ldots, D$, which holds on the condition that the states $|0\rangle,\ldots, |{D-1}\rangle$ are given by Eqs. (\ref{basis-a}) and  (\ref{basis-b}).

A general remark is in order at this place. As follows from Eq. (\ref{rel1}), the average scalar product of two randomly chosen qudit states decreases with growing dimensionality $d$ of the qudits. Hence when $d$ increases, the qudit states get more and more orthogonal on average. It is therefore to be expected that the maximum probability of correct qudit-state identification will also increase with growing dimensionality  $d$. In our paper we shall investigate this dependence quantitatively. 
Since optimal pure-state identification involving unknown states corresponds to the optimal discrimination of mixed states, it is in general hard to find explicit solutions. In the next section we use our method in order to study the optimal identification of qudit states with $d>D$ for those problems where the solution in the case $d=D$ has been already obtained previously.  

\section{Applications}

\subsection{Minimum-error identification}

First we consider the strategy of state identification with minimum error. We apply Eqs. (\ref{Pcorr1}) -  (\ref{minerr1}) and  restrict ourselves to the case where only  two states are to be identified,   occurring with the prior probabilities $\eta_1$ and $\eta_2=1-\eta_1$, respectively. 
Since  ${\Pi}_2^{\Psi}=\hat{P}_{\Psi}- {\Pi}_1^{\Psi}$    the maximum probability of correct results,   $P_{c}^{max} = P_C$, is given by      
\begin{equation}
P_C= \eta_2 - 
  \!\int\! \!{\rm d}\Gamma^{(d)}\left(\eta_2  \langle\Psi_{2}| \Pi_1^{\Psi}  |\Psi_{2}\rangle
-   \eta_1  \langle\Psi_{1}| \Pi_1^{\Psi}  |\Psi_{1}\rangle \right),
\label{Pcorr3}
\end{equation}
where the operator  $\Pi_1^{\Psi}$ maximizes the expression  
\begin{equation}
\label{Pcorr4}
P_c^{\Psi}= \eta_2 -  {\rm Tr}\left[\Lambda_{\Psi}\Pi_1^{\Psi}\right] \quad {\rm with} \;\; \Lambda_{\Psi}=\eta_2 {\rho}_2^{\Psi} - \eta_1 {\rho}_1^{\Psi}. 
\end{equation}
The  spectral decomposition of  $\Lambda_{\Psi}$ can be written as  $\Lambda_{\Psi}=  \sum_{k=0}^{D-1}
\lambda^{(k)} |\pi^{(k)}\rangle \langle \pi^{(k)}|,  $ where we suppose that  $\lambda^{(k)}< 0$ for $k\leq k_0$, while  $\lambda^{(k)}\geq  0$ for all other values of $k$. In  analogy to the solution for the minimum-error discrimination of two mixed states \cite{helstrom,herzog-bergou1}, we obtain the optimal operator    
\begin{equation}
\label{Pi-opt}
\Pi_{1}^{\Psi}= \sum_{k=0}^{k_0}|\pi^{(k)}\rangle \langle \pi^{(k)}| \qquad{\rm if}\;\;  \lambda^{(k)}< 0\;\;{\rm for}\;\; k\leq k_0.
\end{equation}
In other words, the   operator   $\Pi_{1}^{\Psi}$ that maximizes  $P_{c}^{\Psi}$  is equal to the projector onto the subspace  spanned by the eigenstates of $\Lambda_{\Psi}$  belonging to its negative eigenvalues.  

\subsubsection{Minimum-error identification of one known and one unknown  pure qudit state}

We start by assuming that the first qudit state is known, denoted without lack of generality by  $|0\rangle$,  while the second  state, $|\psi\rangle$, is unknown. In this case only a single reference qudit is needed, and Eq. (\ref{Psi}) takes the form   
\begin{equation}
\label{0rho} 
|\Psi_1\rangle =  |0\rangle_0 |\psi\rangle_1   \quad {\rm and} \quad
|\Psi_2\rangle =   |\psi\rangle_0 |\psi\rangle_1.
\end{equation}
  From Eq. (\ref {rho-n-Psi}) with $N=1$ and $N^{\prime}=D=2$ we get the non-normalized density operators 
\begin{equation}
\label{3rho}
\rho_1^{\Psi}= |0\rangle_0\langle 0|_0  \frac {I_1^{(\Psi)}}{d} \quad 
{\rm and}\quad 
\rho_2^{\Psi}= \frac{2}{(d+1)d} \; P_{0,1}^{sym(\Psi)},  
\end{equation}
 which act in the four-dimensional subspace $ {\cal H}_{\Psi} = 
{\cal H}_0^{\Psi}\otimes{\cal H}_1^{\Psi}$. Here ${\cal H}_l^{\Psi}= {\rm span}   \{ |0\rangle_l, |\psi \rangle_l\}$, or, equivalently,   
\begin{equation}
\label{1} 
{\cal H}_l^{\Psi}= {\rm span}   \{ |0\rangle_l, |1 \rangle_l\} \quad {\rm with} \quad   |1\rangle=\frac{|\psi\rangle-  |0\rangle\langle 0|\psi\rangle}{\sqrt{1-|\langle \psi|0\rangle |^2}}.  
\end{equation}
The operators $ I_1^{(\Psi)}$ and $P_{0,1}^{sym(\Psi)}$ follow from  Eqs. (\ref{rel}) and (\ref{P0nsym}) with $d=2$, on the condition that  the state $|1\rangle$ is defined  by Eq.  (\ref{1}).

 Using Eq. (\ref{3rho}) we find that for $\eta_1 > \frac{2}{d+3}$ the operator $\Lambda_{\Psi}= \eta_2 {\rho}_2^{\Psi} - \eta_1 {\rho}_1^{\Psi} $  has exactly two negative eigenvalues, which correspond to the eigenstates   
\begin{equation}
\label{eig} 
|\pi^{(0)}\rangle = c_{1} |0\rangle_0 |1\rangle_1 - c_{2} |1\rangle_0 |0\rangle_1, \quad
|\pi^{(1)}\rangle = |0\rangle_0 |0\rangle_1, 
\end{equation}
where $c_{1/2}\!=\!\sqrt{\frac{1}{2}\pm \frac{(d+1)\eta_1}{2W}}$ with $W\!=\!\sqrt{(d+1)^2\eta_1^2+4 \eta_2^2}$.
 On the other hand,  for $\eta_1 \leq \frac{2}{d+3}$ only the eigenvalue belonging to $|\pi^{(0)}\rangle $  is negative. Because of  Eq. (\ref{Pi-opt}) the optimal operator $\Pi_{1}^{\Psi}$ therefore reads
\begin{equation}
\label{Pi}
\Pi_{1}^{\Psi}=\begin{cases}
|\pi^{(0)}\rangle \langle \pi^{(0)}| &\text{if } \eta_1 \leq \frac{2}{d+3}   \\ 
|\pi^{(0)}\rangle \langle \pi^{(0)}|+|\pi^{(1)}\rangle \langle \pi^{(1)}| &\text{if }  \eta_1 > \frac{2}{d+3}.  
\end{cases}
\end{equation}
In order to apply Eq. (\ref{Pcorr3}) we make use of the expressions $|\langle \Psi_1|\pi^{(0)}\rangle|^2=c_1^2 |\langle \psi|1\rangle |^2$, 
$|\langle \Psi_2|\pi^{(0)}\rangle|^2=(c_1-c_2)^2|\langle \psi|0\rangle |^2|\langle \psi|1\rangle |^2$, 
$|\langle \Psi_1|\pi^{(1)}\rangle|^2=|\langle \psi| 0\rangle|^2$, and $|\langle \Psi_2|\pi^{(1)}\rangle|^2=|\langle \psi| 0\rangle|^4.$ 
Using $|\langle \psi|1\rangle |^2=1-|\langle \psi|0\rangle |^2$ and taking into account that $|0\rangle$ does not depend on $|\psi\rangle,$ the integrations 
in  Eq. (\ref{Pcorr3}) can be easily performed  with the help of Eqs. (\ref{rel1}) and  (\ref{rel4}).  After minor algebra we arrive at the  maximum overall probability of correct results  
\begin{equation}
\label{one-minerr1}
P_C=\frac{d}{d+1}+
\frac{(d-1)(W-d\eta_1)+|2-(d+3)\eta_1|}{2d(d+1)}
\end{equation}
with $W\!=\!\sqrt{(d\!+\!1)^2\eta_1^2+4 (1\!-\!\eta_1^2)}$, see Fig. 1.   When the dimensionality $d$ of the qudits is fixed, $P_C$ takes its smallest value if  $\eta_1= 2/(d+3),$ where $\eta_1$ is the prior probability of the state that is known. In the limit $d\gg 1$  $P_C$ approaches unity, due to the fact that according to  Eq. (\ref{rel1}) in this limit the  modulus of $|\langle 0|\psi \rangle|^2$ tends to zero on average, which means that  the known state $|0\rangle$ gets more and more orthogonal to any other state and perfect discrimination therefore becomes possible.  

We mention that in this simple example it is easy to determine the  optimal operator $\Pi_1$ for discriminating with minimum error between the density operators 
$\rho_1^{(d)}= |0\rangle_0\langle 0|_0  \frac {I_1^{(d)}}{d}$ 
and  
$\rho_2^{(d)}= \frac{2}{(d+1)d} \; P_{0,1}^{sym(d)} ,$ 
see Eq. (\ref{rho_n}), which act in the full  Hilbert space ${\cal H}$. The maximum of the  probability in Eq. (\ref{Pcorr}) is obtained when    
\begin{equation}
\label{Pia}
\Pi_{1}=\begin{cases}
\sum_{j=1}^{d-1}|\pi^{(0)}_{j}\rangle \langle \pi^{(0)}_{j}| &\text{if } \eta_1 \leq \frac{2}{d+3}   \\ 
\sum_{j=1}^{d-1}|\pi^{(0)}_j\rangle \langle \pi^{(0)}_j|+|\pi^{(1)}\rangle \langle \pi^{(1)}| &\text{if }  \eta_1 > \frac{2}{d+3},  
\end{cases}
\end{equation}
 where $|\pi_j^{(0)}\rangle = c_1 |0\rangle_0 |j\rangle_1 - c_2 |j\rangle_0 |0\rangle_1$ and 
where $|\pi^{(1)}\rangle$ is defined  by Eq. $(\ref{eig})$. 
Here the states $|1\rangle, \ldots, |d-1\rangle$ are arbitrary orthonormal basis states in the qudit-subspace that is orthogonal to the known state $|0\rangle$.  By calculating $P_C=  \eta_2- {\rm Tr}[ \Pi_1(\eta_2\rho_2^{(d)}-  \eta_1\rho_1^{(d)}) ]$   Eq. (\ref{one-minerr1}) is regained. In other cases, however, the solution of the  optimization problem in the full Hilbert space ${\cal H}$ may be very cumbersome, or the method to determine the optimal measurement basis  in ${\cal H}$ would not have been even known yet, as  in the last example of our paper, which refers to the optimal unambiguous identification of  $N$ equiprobable unknown qudit states  with $d>N$.

\subsubsection{Minimum-error identification of two unknown pure qudit states}

When we suppose that the two possible pure states of the probe qudit are both unknown, Eq. (\ref{Psi}) yields 
\begin{equation}
\label{Psi1}
|\Psi_{1}\rangle = |\psi_{1}\rangle_0 |\psi_{1}\rangle_1  |\psi_{2}\rangle_2
\quad{\rm and} \quad |\Psi_{2}\rangle = |\psi_{2}\rangle_0 |\psi_{1}\rangle_1  |\psi_{2}\rangle_2.
\end{equation}
From Eq. (\ref{rho-n-Psi}) with  $N=N^{\prime}=D=2$ we get the operators 
\begin{equation}
\label{4rho}
\rho_1^{\Psi}= \frac{2\; P_{0,1}^{sym(\Psi)} I_2^{(\Psi)} }{(d+1)d^2} 
\quad {\rm and}\quad 
\rho_2^{\Psi}= \frac{2\; P_{0,2}^{sym(\Psi)} I_1^{(\Psi)} }{(d+1)d^2}, 
\end{equation}
which  act  
 in the eight-dimensional subspace 
$ {\cal H}_{\Psi} = 
{\cal H}_0^{\Psi}\otimes{\cal H}_1^{\Psi}\otimes{\cal H}_2^{\Psi}$ with
${\cal H}_l^{\Psi}  = 
{\rm span}   \{ |0\rangle_l, |1 \rangle_l\},$  where the states $|0\rangle$ and $  |1 \rangle$ are given by  Eq. (\ref{basis-a}). 
\begin{figure}
\center{\includegraphics[width=8.5 cm]{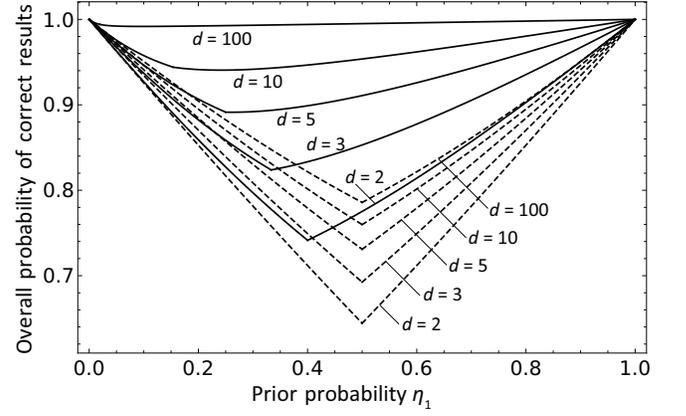}}
 \caption {
Maximum overall probability $P_C$ of correct results for the minimum-error identification of two qudit states when the first state is known and the second is unknown [full lines, corresponding to Eq. (\ref{one-minerr1})] and when both states are unknown  [dashed lines, corresponding to Eq. (\ref{P_C})]    
vs the prior probability $\eta_1$ of the first state,  for different values of the qudit dimensionality $d$.}
\end{figure}

To be specific, we  first assume that  $\eta_1 \leq \eta_2$. In this case only two eigenvalues of  the operator $\Lambda_{\Psi}=\eta_2 {\rho}_2^{\Psi} - \eta_1 {\rho}_1^{\Psi}$ are negative.  After calculating the corresponding eigenstates and denoting them again  by $|\pi^{(0)}\rangle $ and $|\pi^{(1)}\rangle$,  we obtain from Eq. (\ref{Pi-opt})  the optimal operator    
\begin{equation}
\label{min0} 
\Pi_{1}^{\Psi}=\sum_{k=0}^1|\pi^{(k)}\rangle \langle \pi^{(k)}|
\quad {\rm for}\;\;\eta_1 \leq \eta_2, \quad {\rm where}  
\qquad\qquad
\end{equation}
\begin{equation}
\label{min1}
|\pi^{(0)}\rangle =\frac{|1\rangle_0 |0\rangle_1 |0\rangle_2- (1+b)|0\rangle_0 |0\rangle_1 |1\rangle_2  +b |0\rangle_0 |1\rangle_1 |0\rangle_2}  {\sqrt{2(1+b+b^2)}},
\end{equation}
\begin{equation}
\label{min2}
|\pi^{(1)}\rangle  =\frac{|0\rangle_0 |1\rangle_1 |1\rangle_2- (1+b)|1\rangle_0 |1\rangle_1 |0\rangle_2  +b |1\rangle_0 |0\rangle_1 |1\rangle_2}  {\sqrt{2(1+b+b^2)}}\,
\end{equation}
with $b=\sqrt{1-\eta_1\eta_2}-\eta_2$.  
Applying Eq. (\ref{Pcorr3})  we find that   
\begin{equation}
 P_{C}= \eta_2 - \frac{\eta_2\, b^2-\eta_1 (1+b)^2 } {2(1+b+b^2)} \!  \int\! \!{\rm d}\Gamma_2^{(d)}  |\langle\psi_2|1\rangle|^2
\quad {\rm if}\;\;\eta_1 \leq \eta_2,
\label{Pcorr5}
\end{equation}
where we took into account that    $\langle \psi_1|0\rangle =1,$  $\langle \psi_1|1\rangle =0$ and 
  $|\langle\psi_2|1\rangle|^2 |\langle\psi_2|0\rangle|^2 +  |\langle\psi_2|1\rangle|^4=|\langle\psi_2|1\rangle|^2,$ as follows from Eq. (\ref{basis-a}). 
A similar expression for $P_C$ holds  in the case $\eta_2 \leq \eta_1$.  
Using $ \int \!{\rm d}\Gamma_2^{(d)} |\langle\psi_2|1\rangle|^2=1-\frac{1}{d}$, see  Eq. (\ref{rel2}),  
 after minor algebra  we arrive at  the maximum overall probability of correct identification 
\begin{equation}
\label{P_C}
P_{C}=\frac{1}{2}+\frac{d+2}{6d}|\eta_2-\eta_1|+\frac{d-1}{3d}\sqrt{1-\eta_1\eta_2},
\end{equation}
in accordance with the result that was obtained in Ref. \cite{IHHH} using a different approach. As becomes obvious from Fig. 1, $P_C$ takes its smallest value when the two unknown states are equiprobable. In this case we get $P_C= \frac{1}{2}+ \frac{d-1}{2d\sqrt{3}}$, which yields $P_C \approx 0.79$ in the limit  $d \gg 1$.

\subsection{Optimal unambiguous identification}

\subsubsection{Optimal unambiguous identification of one known and one unknown pure qudit state}

Now we turn to the strategy of optimal unambiguous state identification. Because of  Eqs. (\ref{Pcorr1}), (\ref{Pcorr2}), and  (\ref{unamb1}) with $N^{\prime}=2$,
the optimal  operators  $\Pi_1^{\Psi}$  and  $\Pi_2^{\Psi}$ have to maximize the expression  $P_c^{\Psi}= \sum_{n=1}^2\eta_n {\rm Tr}\left(\rho_n ^{\Psi}\Pi_n^{\Psi}\right)$  on the condition that  
\begin{equation}
\Pi_0^{\Psi}=\hat{P}_{\Psi}- \Pi_1^{\Psi}-\Pi_2^{\Psi} \geq 0, \quad  \Pi_1^{\Psi}\rho_2^{\Psi}=  \Pi_2^{\Psi}\rho_1^{\Psi}=0
\label{unamb2}.
\end{equation}
 In contrast to minimum-error discrimination, instead of a general solution only an upper bound has been obtained for the maximum success probability when two arbitrary mixed states are to be unambiguously discriminated \cite{herzog-bergou2}.  However, due to the special structure of $ \rho_1^{\Psi}$ and 
$\rho_2^{\Psi}$  the maximization of $P_c^{\Psi}$ can be easily performed in our case.  

Let us again consider the situation   where  the first state is the known state $|0\rangle$ and the second  state is unknown, as described by Eqs. (\ref{0rho}) - (\ref{1}).
In analogy to our earlier treatment for $d=2$ \cite{BBFHH} we find that  due to the constraints  given by  Eq. (\ref{unamb2})  the optimal operators  
take the general form    
\begin{eqnarray}
\label{one-unamb2}
\!\!\!\!\!\!\!&&\Pi_1^{\Psi}\!= \alpha_1 |\pi_1\rangle \langle \pi_1|\quad{\rm with}\;\;
                       |\pi_1\rangle=\frac{|0\rangle_0 |1\rangle_1- |1\rangle_0 |0\rangle_1 } {\sqrt{2}},\;\qquad\;\\
\label{one-unamb3}
\!\!\!\!\!\!\!&&\Pi_2^{\Psi}\!=\alpha_2 |1\rangle_0 |0\rangle_1 \langle 1|_0 \langle 0|_1
+|1\rangle_0 |1\rangle_1 \langle 1|_0 \langle 1|_1,
\end{eqnarray}
where $\alpha_2=\frac{2-2 \alpha_1}{2- \alpha_1},$ which is the largest value of $\alpha_2$ compatible with the constraint  $\Pi_0^{\Psi}\geq 0.$ Using Eq. (\ref{3rho}) and  maximizing   $P_{c}^{\Psi} $ in Eq. (\ref{Pcorr2}), we obtain the optimal parameter 
\begin{equation}
\label{one-unamb4}
\alpha_1=\begin{cases}
0 &\text{if } (d+1)\eta_1 \leq \eta_2    \\ 
2-2\sqrt{\frac{\eta_2}{(d+1)\eta_1}} &\text{if }\eta_2 \leq  (d+1)\eta_1   \leq  4\eta_2  \\
 1 &\text{if } (d+1)\eta_1 \geq 4\eta_2.   
\end{cases}
\end{equation}
Here the  requirement $0\leq \alpha_1\leq 1$ has been taken into account,  which  results from the positivity of the optimal operators and from the completeness relation in the form of Eq. (\ref{pos}). 
With the help of  Eqs.  (\ref{Pcorr1}) and (\ref{0rho}) the optimal operators $\Pi_1^{\Psi}$ and  $\Pi_2^{\Psi}$ yield   
$P_{c}^{max} = P_S,$ where 
\begin{eqnarray}
\label{one-unamb5}
\!\!\! P_S= \int \!{\rm d}\Gamma^{(d)}\left(\eta_1\frac{\alpha_1}{2}|\langle \psi|1\rangle|^2  \right. \quad\quad \qquad\qquad\qquad\qquad \\
   \left. +\eta_2\alpha_2|\langle \psi|1\rangle|^2|\langle \psi|0\rangle|^2  +\eta_2 |\langle \psi|1\rangle|^4\right). \nonumber
\end{eqnarray}
 Inserting  $ |\langle \psi|1\rangle|^2=1-|\langle \psi|0\rangle|^2$ and taking into account that $|0\rangle$ does not depend on $|\psi\rangle,$ the integrations over the state space of the unknown state $|\psi\rangle$ can be easily performed by means of   Eqs. (\ref{rel1}) and (\ref{rel4}). 
Using $\eta_2=1-\eta_1,$ we finally arrive at the maximum overall success probability for unambiguous identification  
\begin{equation}
\label{one-unamb6}
P_S =\begin{cases}
\frac{d-1}{d}(1-\eta_1)  &\text{if } \eta_1\leq \frac{1}{d+2}    \\ 
\frac{d-1}{d}\!\left(\!\frac{d+2-\eta_1}{d+1}-2\sqrt{\frac{\eta_1(1-\eta_1)}{d+1}}\!\right )
 &\text{if }  \frac{1}{d+2}\leq \eta_1 \leq \frac{4}{d+5}\\
\frac{d-1}{d+1}\left(1-\eta_1\frac{d-1}{2d}\right) 
&\text{if }  \eta_1\geq \frac{4}{d+5},
\end{cases}
\end{equation}
see Fig. 2.
The upper full line in Fig. 2 results from $\alpha_1=0$ and $\alpha_2=1$, which means that $\Pi_1^{\Psi}=0$ and $\Pi_2^{\Psi}=  |1\rangle_0\langle 1|_0\otimes I_1^{\psi}.$ In this case the known state is never identified but always yields an inconclusive result, and the unambiguous identification of the second state is realized when a  projection onto a state orthogonal to the known state $|0\rangle_0$ is successful. 
On the other hand, if  $\alpha_1\neq  0,$ that is in the parameter region where the middle and the lower line of Eq. (\ref{one-unamb6}) apply, the operators $\Pi_1^{\Psi}$ and $\Pi_2^{\Psi}$ are both different from zero. 
The lower line  of Eq. (\ref{one-unamb6}) is valid  when  $\alpha_1=1$ and  $\alpha_2=0$, that is when  $\Pi_1^{\Psi}= |\pi_1\rangle \langle \pi_1| $ and  
$\Pi_2^{\Psi}= |1\rangle_0 |1\rangle_1 \langle 1|_0 \langle 1|_1$. 
In the limit $\eta_1 \rightarrow 1$  the success probability $P_S$ 
is only determined by  $\Pi_1^{\Psi}$, which according to Eq. (\ref{one-unamb2}) with $\alpha_1=1$ then represents a projection onto the antisymmetric subspace of the joint Hilbert space $ {\cal H}_{\Psi}$
belonging to the probe qudit and the reference qudit encoding  the second state.

\subsubsection{Optimal unambiguous identification of two unknown pure qudit states} 

\begin{figure}[t!]
\center{\includegraphics[width=8.5 cm]{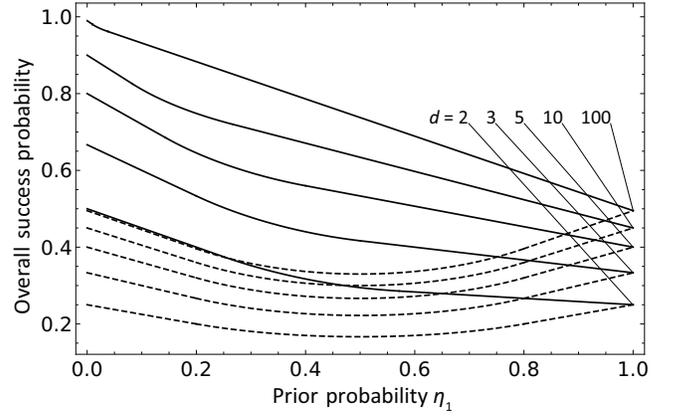}}
 \caption {Maximum overall success  probability $P_S$ for the  unambiguous identification of two qudit states when the first state is known and the second is unknown  [full lines, corresponding to Eq. (\ref{one-unamb6})] and when both states are unknown [dashed lines, corresponding to Eq. (\ref{two-unamb2})]     
vs the prior probability $\eta_1$  of the first state, for different values of the qudit dimensionality $d$. }
\end{figure}
When both qudit states are unknown, we start from Eqs. (\ref{Psi1}) and (\ref{4rho}).  The  constraints in  Eq. (\ref{unamb2})  then imply that the projector onto the support of $\Pi_1^{\psi}$  is given by $I_1^{(\Psi)} P_{0,2}^{as(\Psi)}\!$,  in analogy to the treatment for $d=2$ \cite{BH,BBFHH}. Here  $P_{0,2}^{as(\Psi) }= \hat{P}_{\Psi}-P_{0,2}^{sym(\Psi) }$ 
 projects onto the antisymmetric state  $(|0\rangle_0|1\rangle_2 - |1\rangle_0|0\rangle_2) / \sqrt{2}$, where   $|0\rangle$ and $  |1 \rangle$ are given by  Eq. (\ref{basis-a}).  A similar expression holds for the projector onto the support of $\Pi_2^{\psi}$. 
The optimal operators  
take the  form 
\begin{eqnarray}
\label{Phi-two}
\Pi_n^{\Psi}= \alpha_n \sum_{k=0}^1|\pi^{(k)}_n\rangle \langle \pi^{(k)}_n| \quad(n,k=1,2)\qquad\qquad\\
{\rm with}
\;\;|\pi_1^{(k)}\rangle = \frac{1}{\sqrt{2}}|k\rangle_1\left (|0\rangle_0|1\rangle_2 - |1\rangle_0|0\rangle_2\right)\qquad\; \\
\;\;|\pi_2^{(k)}\rangle = \frac{1}{\sqrt{2}}|k\rangle_2\left (|0\rangle_0|1\rangle_1 - |1\rangle_0|0\rangle_1\right),\qquad
\end{eqnarray}
where $\alpha_2=\frac{4-4 \alpha_1}{4- 3\alpha_1}$, which is the largest value of $\alpha_2$ compatible with the constraint  $\Pi_0^{\Psi}\geq 0$. The explicit values of $\alpha_1$ and $\alpha_2$, arising from  
the maximization of  $P_{c}^{\Psi}$  in Eq. (\ref{Pcorr2}), are the same as the values that have been obtained for $d=2$ \cite{BH,BBFHH}. 
Using the resulting optimal operators $\Pi_1^{\Psi}$ and $\Pi_2^{\Psi},$ together with the  expressions for $|\Psi_1\rangle$ and $|\Psi_2\rangle$ given by Eq. (\ref{Psi1}),  the  integration in  Eq. (\ref{Pcorr1}) can be easily carried out with the help of  Eqs. (\ref{basis-a}) and  (\ref{rel2}). We thus obtain the maximum overall success probability  $P_{c}^{max} = P_S,$ which takes the form 
\begin{equation}
\label{two-unamb2}
P_S =\begin{cases}
\frac{d-1}{2d}(1-\eta_1)  &\text{if } \eta_1\leq \frac{1}{5}    \\ 
\frac{2(d-1)}{3d}\left(1- \sqrt{\eta_1(1-\eta_1)}\right)
 &\text{if } \frac{1}{5} \leq \eta_1 \leq \frac{4}{5}\\
\frac{d-1}{2d}\eta_1 &\text{if }  \eta_1\geq \frac{4}{5}.
\end{cases}
\end{equation}
The upper line refers to the parameter region where $\alpha_1=0$ and $\alpha_2=1$. Vice versa, the lower line applies when $\alpha_2=0$ and $\alpha_1=1,$ that is when $\Pi_2^{\Psi} =0$  and  $\Pi_1^{\psi}=I_1^{(\Psi)} P_{0,2}^{as(\Psi) },$ which means  that  the action of  $\Pi_1^{\psi}$ corresponds to a projection onto the antisymmetric subspace of the joint Hilbert space  ${\cal H}_0^{\Psi}\otimes{\cal H}_2^{\Psi}$ belonging to the probe qudit and the reference qudit encoding the second state.  In the limit $\eta_1 \rightarrow 1,$  where $P_S$ is only determined by $\Pi_1^{\psi}$,  the  maximum success probabilities resulting from Eqs. (\ref{one-unamb6}) and (\ref{two-unamb2}) are identical, see Fig. 2.  

We still mention that in Ref. \cite{BH} the optimal measurement for the unambiguous identification of two unknown pure states has been applied to two fixed states $|\psi_1\rangle$ and  $|\psi_2\rangle$, yielding expressions for $P_S$  in the three regions of  $\eta_1$ that each contain  a factor $(1-|\langle \psi_1|\psi_2\rangle|^2)$  \cite{BH}.  When we assume that the states are qudits and use  Eq. (\ref{rel1}) in order to take  the average in this factor, Eq. (\ref{two-unamb2}) is regained. 

\subsubsection{Optimal unambiguous identification of $N$  
linearly independent  and equiprobable unknown pure  qudit states}

In our last example we apply the general results derived in Sec. II to the optimal unambiguous identification of $N$ linearly independent unknown  pure qudit 
states with dimensionality $d\geq N.$
 Since the $N$ states are linearly independent, they span a Hilbert space of dimension $D=N$.
 We restrict ourselves to the case of equiprobable states, occurring  with equal prior probabilities $\eta_n=1/N$. 
Due to  Eq.  (\ref{Pcorr1}) with $N^{\prime}=N$,  the overall probability of correct results  then can be written as    
\begin{equation}
P_c
=   \frac{1}{N} \sum_{n=1}^{N} \!\int\! \!{\rm d}\Gamma^{(d)}   \langle\Psi_{n}| \Pi_n^{\Psi}  |\Psi_{n}\rangle,
\label{PcorrN}
\end{equation}
where $|\Psi_{n}\rangle = |\psi_{n}\rangle_0 \;|\psi_{1}\rangle_1 \ldots  |\psi_{N}\rangle_N. $ The operators $\Pi_n^{\Psi}$  act in the $N^{N+1}$-dimensional Hilbert space ${\cal H}_{\Psi}$ defined in Eq. (\ref{H-psi}), which is a subspace of the total $d^{N+1}$-dimensional Hilbert space  ${\cal H}.$ 
 In order to find the optimal operators $\Pi_n^{\Psi}$  we have to maximize 
\begin{equation}
\label{N1a}
P_c^{\Psi}=  
\frac{1}{N}\sum_{n=1}^{N} {\rm Tr}\left(\rho_n ^{\Psi}\Pi_n^{\Psi}\right), 
\end{equation}
as follows from Eq. (\ref{Pcorr2}), where the maximization is subject to the constraints given by Eq. (\ref{unamb1}). 
According to  Eq. (\ref{rho-n-Psi}) the  non-normalized  density operators $\rho_n^{\Psi}$ 
take the form
\begin{equation}
\label{N1}
\rho_n^{\Psi}= \dfrac{2}{(d+1)d^N}\;P_{0,n}^{sym(\Psi)} \bigotimes \limits_{l=1\atop
l\neq n}^N I_l^{(\Psi)} \quad (n=1,\ldots,N). 
\end{equation}
Here the operators $ I_l^{(\Psi)}$ and $P_{0,n}^{sym(\Psi)}$ are determined  by  Eqs. (\ref{rel}) and (\ref{P0nsym}) with $d=N,$ where  the basis   states $|0\rangle_l,\ldots, |N-1\rangle_l$ are defined  by Eqs.  (\ref{basis-a}) -  (\ref{basis-c}).
 The resulting maximization problem 
 is mathematically equivalent to the problem we solved already in our previous work \cite{HB-d}, where we considered the optimal unambiguous identification of $d$ linearly independent unknown pure qudit states,  that is the case $d=N$. Using our previous result \cite{HB-d}, we get the representation  
\begin{eqnarray}
\label{N4}
&&\Pi_n ^{\Psi} = \frac{N}{N+1}\sum_{k=0}^{N-1} |\pi^{(k)}_n\rangle \langle \pi^{(k)}_n|\\
&&{\rm with}\quad|\pi^{(k)}_n\rangle  = \frac{(-1)^{n}}{\sqrt{N!}}|k\rangle_n
\sum_{\sigma}
{\rm sgn} (\sigma)\! \bigotimes_{j=0\atop j\neq
n}^N|{\sigma_j}\rangle_j,
 \label{pi-n}\qquad\quad
\end{eqnarray}
 where the sum in the second line is taken over all $N!$ permutations
$\sigma$ distributing the numbers $\sigma_{j}=0,\ldots, N-1$
over the system of $N$ qudits  obtained by omitting the $n$-th
reference qudit from the total system of $N+1$ qudits. The
qudits are written in fixed order, and ${\rm sgn} (\sigma)$ is the
sign of the permutation. For instance, for $N=3$ we get  
\begin{eqnarray}
|\pi^{(k)}_1\rangle \!= \!\frac{-1}{\sqrt{6}} |k\rangle_1 \Big(
|0\rangle_0|1\rangle_2|2\rangle_3
-|0\rangle_0|2\rangle_2|1\rangle_3 +|2\rangle_0|0\rangle_2|1\rangle_3 \nonumber \\
- |2\rangle_0|1\rangle_2|0\rangle_3  
+|1\rangle_0|2\rangle_2|0\rangle_3-|1\rangle_0|0\rangle_2|2\rangle_3
  \Big) .
\qquad
 \label{N5}
\end{eqnarray}
It is easy to check that $P_{0,m}^{sym(\Psi)}|\pi^{(k)}_n\rangle =0$ if $ m \neq n, $ which ensures  that the constraint 
${\rho_m^{\Psi}}{\Pi}_n^{\Psi}  
=0$,  required for unambiguous identification, is satisfied. The prefactor $N/(N+1)$ in Eq. (\ref{N4}) yields the largest possible value of $P_c^{\Psi}$ that is compatible with the positivity  constraint $ \Pi_0^{\Psi}=\hat{P}_{\Psi}- \sum_{n=1}^N \Pi_n^{\Psi} \geq 0$, as has been shown in Ref. \cite{HB-d} by using the relations
$\langle \pi^{(k)}_{n} |\pi^{(k^{\prime})}_n\rangle = \delta_{k,k^{\prime}}$ and $\langle \pi_m^{(k)}|\pi^{(k^{\prime})}_n\rangle
 = - \frac{ \delta_{k,k^{\prime}}}{N} \;\; (m\neq n)$.
\begin{figure}[t!]
\center{\includegraphics[width=8.5 cm]{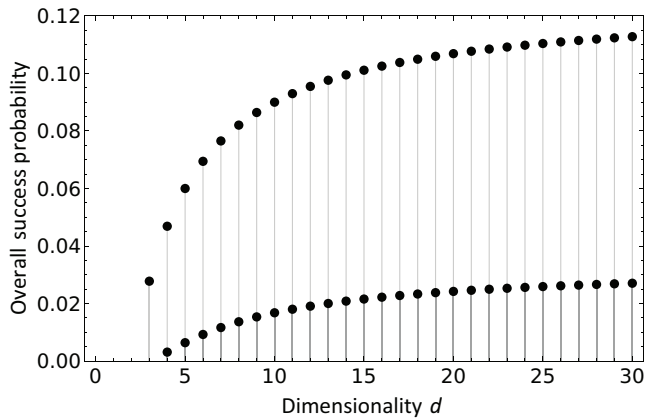}}
 \caption {Maximum overall success  probability $P_S$ for the unambiguous identification of $N$ equiprobable linearly independent  unknown pure qudit states versus the dimensionality $d$ of the qudits for $N=3$ (upper dots) and $N=4$ (lower dots), as given by Eq. (\ref{PcorrN2}).     }
\end{figure}

The values of ${\rm Tr}\left(\rho_n ^{\Psi}\Pi_n^{\Psi}\right)$ resulting from Eqs. (\ref{N1}) and (\ref{N4}) are found to be the same for each  number $n$, which is due to the fact that the optimal operators $\Pi_n ^{\Psi}$  obey the same permutation symmetry with respect to numbering the states as the density operators $\rho_n^{\Psi}.$  
Instead of Eq. (\ref{N1a}) we can therefore use the alternative expression  $P_c^{\Psi}={\rm Tr}\left(\rho_1 ^{\Psi}\Pi_1^{\Psi}\right)$ for the maximum of  $ P_c^{\Psi}$. Similarly,  
due to the permutation symmetry it follows from Eqs. (\ref{PcorrN}) and  (\ref{N4})  that the  maximum success probability  $P_S = P_{c}^{max}$ can be written as 
\begin{equation}
 P_S\!=\! 
   \!\int\! \!{\rm d}\Gamma^{(d)}  \langle\Psi_{1}| \Pi_1^{\Psi}  |\Psi_{1}\rangle\!=\!\frac{N}{N\!+\!1}\!\sum_{k=0}^{N-1} \!\int\! \!{\rm d}\Gamma^{(d)}|\langle \Psi_1|\pi^{(k)}_{1}\rangle |^2.
\label{PcorrN1}
\end{equation}
Inserting $|\Psi_{1}\rangle = |\psi_{1}\rangle_0 \;|\psi_{1}\rangle_1 \ldots  |\psi_{N}\rangle_N$ and taking  into account that in the expression for $|\pi^{(k)}_{1}\rangle $ the  states $|0\rangle,\ldots, |N-1\rangle$  are defined  by Eqs.  (\ref{basis-a}) -  (\ref{basis-c}), we find that 
\begin{equation}
\label{N6}
\int\! \!{\rm d}\Gamma^{(d)}  |\langle \Psi_1|\pi^{(k)}_{1}\rangle |^2
=  \frac{\delta_{k,1}}{{N!}}  \prod_{j=2}^N \int \!{\rm d}\Gamma_j^{(d)}  |\langle \psi_j|j-1\rangle_j|^2.
\end{equation}
 Here we made use of the relations $\langle \psi_1|k-1 \rangle= \delta_{k,1} $ and 
 $\langle \psi_j|k \rangle = 0$ for $k \geq j$, see Eq. (\ref{basis1}).  The latter condition implies that only a single one from the $N!$ terms occurring in the expression for $|\pi_1^{(1)}\rangle$  yields a non-zero contribution to the scalar product   $\langle \Psi_1|\pi_1^{(1)}\rangle,$ as can be easily exemplified for $N=3.$    
After applying Eq. (\ref{rel2}) in order to perform the integrations in Eq. (\ref{N6}),  we obtain for $d\geq N$  the maximum overall success probability 
\begin{equation}
\label{PcorrN2}
P_S= 
\frac{N \left(1-\frac{1}{d}\right)\ldots \left(1-\frac{N-1}{d}\right)} {(N+1)\, N!}   =\frac{N}{(N+1)\,d^N}{d \choose N}.
\end{equation}
For a fixed number $N$ of unknown qudit states the overall probability of successful identification, $P_S$, increases with growing dimensionality $d$ of the qudits, see Fig. 3. From Eq. (\ref{PcorrN2}) it becomes obvious that  $P_S=   \frac{N}{N+1}\frac{1}
{N^N}$  for $d=N$ \cite{HB-d} and  $P_S=   \frac{N}{N+1}\frac{1}
{N!} $ for  $d\gg N.$ 
Hence by sufficiently increasing the dimensionality $d$ when $N$ is fixed,  $P_S$ can be enhanced by a factor  up to ${N^N}/{N!}$. Although this factor      
grows with increasing $N$, the absolute value of $P_S$ is rapidly decreasing when the 
number of states is enlarged.  

\section{Conclusions}

In this paper we considered the optimal identification of pure qudit states with dimensionality $d$ that belong to a finite set of states where some or all of the states  are unknown. 
Supposing that  the  qudit states span a $D$-dimensional Hilbert space, we found that from an optimal measurement for state identification with   $d=D$ one can  readily determine the optimal figure of merit for qudits with  $d>D,$ without solving a new optimization problem, that is without explicitly determining the optimal detection operators acting in the full Hilbert space.    
We applied our method to a  number of examples, including the optimal unambiguous identification of $N$  
linearly independent  and equiprobable unknown pure  qudit states. In all cases we found that the maximum overall probability of correct results for minimum-error identification, or the maximum overall success probability for unambiguous identification, respectively,  increase with growing dimensionality $d$ of the qudits. 

The results of this paper may be of interest when high-dimensional quantum states  are to be processed,  which have been considered as a resource for various tasks in quantum information and communication \cite{goyal, chau, bellomo}. These states can  be for instance produced as superpositions of orbital angular momentum states of a single photon, and  the optimal unambiguous discrimination of $d$ linearly independent symmetric pure qudit states has been already experimentally realized for known states with dimensions up to  $d=14$ \cite{agnew}. 
While the problems of comparing \cite{BCJ} and of identifying unknown quantum states have been introduced already about a decade ago,  
very recently also superpositions of unknown states have been  investigated \cite{oszmaniec}.

\begin{acknowledgments}
The author is grateful for many useful discussions with J\'{a}nos Bergou on various aspects of state discrimination.
\end{acknowledgments}

\end{document}